\documentclass[10pt,twocolumn]{article}
\usepackage{subfigure}
\usepackage{epsfig}
\usepackage{graphicx}
\usepackage{layout}
\usepackage{multicol}
\usepackage{times}
\date{}
\graphicspath{{image/}}
\begin{document}
\twocolumn[ 
\begin{center}
\textsf{\LARGE \textbf{Framework and Bio-Mechanical Model for a Per-Operative Image-Guided Neuronavigator Including `Brain-Shift' Compensation}}
\end{center}
\textsf{
\begin{multicols}{2}
\begin{center}
{ \large Marek Bucki }$^{1,2}$\\
{ \large Yohan Payan }$^1$\\
\end{center}
\end{multicols}
}
$^1$ TIMC Laboratory, UMR CNRS 5525, University J. Fourier, 38706 La Tronche, France \\
$^2$ Centro de Modelamiento Matematico, Av. Blanco Encalada 2120, Santiago de Chile, Chile
\begin{center}
{\large mbucki@dim.uchile.cl }
\end{center}


\begin{center}
{\large \textbf{ABSTRACT} }
\end{center}
In this paper we present a methodology to adress the problem of brain tissue deformation referred to as "brain-shift". This deformation occurs throughout a neurosurgery  intervention and strongly alters the accuracy of the neuronavigation systems used to date in clinical routine which rely solely on preoperative patient imaging to locate the surgical target, such as a tumour or a functional area. After a general description of the framework of our intraoperative image-guided system, we propose a  biomechanical model of the brain which can take into account interactively such deformations as well as surgical procedures that modify the brain structure, like tumour or tissue resection.
\vspace{6pt}

\vspace{14pt}
]

\section{INTRODUCTION}
Accurate localization of the target is essential to reduce the morbidity during a brain tumor removal intervention. Image-guided neurosurgery is facing an important issue for large skull openings, with intraoperative changes that remain largely unsolved. In that case, deformations of the brain tissues occur in the course of surgery because of physical and physiological phenomena. As a consequence of this brain-shift, the preoperatively acquired images no longer correspond to reality; the preoperative based neuronavigation is therefore strongly compromised by intraoperative brain deformations. Some studies have tried to measure this intra-operative brain-shift. Hastreiter et al. \cite{Hastreiter04} observed a great variability of the brain-shift ranging up to 24 mm for cortical displacement and exceeding 3 mm for the deep tumor margin; the authors claim for a non-correlation of the brain surface and the deeper structures. Nabavi et al. \cite{Nabavi01} state that the continuous dynamic brain-shift process evolves differently in distinct brain regions, with a surface shift that occurs throughout surgery (and that the authors attribute to gravity) and with a subsurface shift that mainly occurs during resection (that the authors attribute to the collapse of the resection cavity and to the intraparenchymal changes). In order to face this problem, authors have proposed to add to actual image-guided neurosurgical systems a module to compensate brain deformations by updating the preoperative images and planning according to intraoperative brain shape changes. The first algorithms developed proposed to deform the preoperatively acquired images using image-based models. Different non-rigid registration methods were therefore provided to match intraoperative images (mainly MRI exams) with preoperative ones \cite{Hata00}, \cite{Hastreiter00}, \cite{Shattuck02}. More recently, biomechanical models of the brain tissues were proposed to constrain the image registration: the models are used to infer a volumetric deformation field from correspondences between contours \cite{Kyriacou99}, \cite{Hagemann99} and/or surfaces \cite{Ferrant02} in the images to register. Arguing against the exorbitant cost of the intraoperative MRI imaging devices, some authors have proposed to couple the biomechanical model of the brain with low-cost readily available intraoperative data \cite{Miga05} such as laser-range scanner systems \cite{Audette03}, \cite{Miga03} or intraoperative ultrasound \cite{Comeau00}. This proposal seems appealing from a very practical point of view, compared with the high cost intraoperative MRI device. However, it gives to the biomechanical model a crucial and very central position. This means that a strong modelling effort has to be carried out during the design of the brain biomechanical model as well as its validation through clinical data.

In section \ref{SecGeneralFramework} we present the framework we have chosen to tackle the brain-shift problem. In section \ref{SecImplementation} we present an overview of the implementation of the bio-mechanical model. Then an example of 2D brain-shift simulation is presented in section \ref{SecExampleOfSimulation}, followed by the conclusion.

\section{GENERAL FRAMEWORK}
\label{SecGeneralFramework}
During a tumor resection the goal of the intervention is to minimize the injury while maximizing resection. In case of a large brain-shift, without a computerized assistance and relying only on the vision of the operating field provided by preoperative imaging, it is a difficult task for the surgeon to preserve eloquent cortex areas and white matter tracts which are subject to shift at various stages of the procedure as shown by Coenen et al. in \cite{Coenen05}. In order to be applicable to a large spectrum of interventions, a neuronavigation system must comprise a precise brain-shift compensation module.

The neuronavigation system we are currently working on relies on the following ideas. (1) A finite-element biomechanical model of the brain driven by the position of anatomical landmarks is used to compute the global tissue deformation. (2) The deformation is applied to operating room(OR)-registered preoperative patient data in order to predict the location of the target. (3) The deformation occurs throughout the intervention and needs to be monitored as often as possible. (4) The deformation is due to multiple factors, some of which are hard to monitor like tissue inflamation and CSF leakage. Others can be integrated within the system like gravity effect, tissue resection, cyst drainage and deep structures displacements. We assume that the information provided by the latter is sufficient to account precisely for the global deformation

The goals of our system are to predict the initial brain-shift that occurs right after the dura opening; to follow the brain deformation throughout the surgery using per-operative data; to account for tissue modification such as tumor resection with low computational delays; and to rely mainly on low-cost imaging techniques such as localised 2.5D echography for deep structures displacements while a solution such as the one presented by Miga et al. in \cite{Miga03} could track cortical displacements.


\subsection{Biomechancial model}
\label{SubsecBiomechanicalAssumptions}
Miller has shown in his works \cite{Miller02} that brain tissue has a visco-elastic, non-linear behaviour which yields equations difficult to integrate in real-time. A simpler model is thus necessary, especially if tissue resection needs to be modeled. Picinbono et al. proposed  in \cite{Picinbono02} a large-deformation mass-tensor approach but we are not sure that large displacements are really observed in the case of brain-shift. To our knowledge, Clatz et al. mention $7\%$ \cite{Clatz05}\cite{Clatz05-RR} deformations in the case of Parkinson interventions with small apertures and we didn't find any results showing higher order deformation ($10$ or $15\%$). We thus chose, as a first approximation, a mechanical linear and small deformations model which will require validation against clinical data. This hypothesis allows us to model in an interactive way, surgical interactions such as cyst drainage and tissue resection, as will be described below.

We assume slow deformations of the parenchyma and thus consider the static equilibrium equation $div(\sigma)+F_{ext} = 0$ and use the linear expression of strain and stress tensor, respectively $\epsilon = \frac{1}{2}(\nabla U+\nabla U^t)$ and $\sigma = C\epsilon$, where $U$ is the displacement field and $C$ the material matrix.


\subsection{Behaviour hypothesis}
\label{SubsecBehaviourHypothesis}
We assume that initial brain bulging observed at the time of dura opening is due to intracranial pressure increase generated by the tumor growth. The tumor compresses the brain structures such as ventricules, falx cerebri and the contralateral hemisphere which is clearly observable on the MRI - see Figure \ref{FigSimulation}-A: the dashed line is the interface between the two hemispheres; the dotted line is the initial position of this interface. When the dura is opened, the available space creates a new equilibrium position within the brain, generating the initial brain shift.

We predict the amplitude of this bulging my modeling the contralateral hemisphere, computing the stress generated by the compression and applying this constraints on the hemisphere with tumor. The limit conditions in the contralateral model are (1) inter-hemispheric limit displacement applied to border nodes, and (2) non-friction sliding of the border nodes in contact with the skull. The limit conditions in the tumor model are (1) initial constraints at the inter-hemispheric limit computed from contralateral model deformation, (2) non-friction sliding of the border nodes in contact with the skulll, except near the craniotomy, where the nodes can move freely.

The mecanical parameters mentionned in paragraph \ref{SubsecBiomechanicalAssumptions} can only be valid if the brain tissue compression induced by tumor growth is fast and does not let the tissue loose its elasticity by long exposition to permanent strain. Fast evolving tumors such as gliomas might be described by this model.

We assume that brain sagging is mainly due to gravity associated to CSF leakage. The other factors involved in brain sagging are tissue resection and tumoral cyst drainage. The modeling of these intraoperative changes will be described in section \ref{SecImplementation}.


\subsection{Preoperative process}
The tumor is first identified on the preoperative images by the radiologist on the MRI scan and the two brain hemispheres are segmented separately. The system uses this processed images to construct a 3D mesh of the brain that will be used for finite elements modeling. The mesh construction algorithm is currently being developped and relies on the surface reconstruction algorithm presented by the authors in \cite{Bucki05}.

The surgeon then defines the approach route, localising as well the craniotomy site. In the 2 meshes the nodes are labelled as `pilot nodes' i.e. nodes associated with reference structures and which displacement can be updated using intraoperative imaging; `sliding nodes' i.e. nodes in contact with the skull which displacement is constrained along the tangent line to the skull in 2D, and tangent plane in 3D; `free nodes' i.e. nodes without any displacement constraint. The contralateral deformation can then be computed and the initial brain bulging estimated as described in \ref{SubsecBehaviourHypothesis}.


\subsection{Intraoperative process}
First the data are rigidly registered to the OR patient configuration. Then, at regular time intervals echographic sparse data is segmented in order to update the biomechanical brain model based on reference structure displacements. Once the tumor-brain interface has been prepared, the tissue excision begins. During this phase, the surgical tools are localised in space and tissue volume resection is evaluated in order to apply the topological changes induced in the brain mesh by the resection or cyst drainage.




\section{IMPLEMENTATION}
\label{SecImplementation}
The linear mechanics described in \ref{SubsecBiomechanicalAssumptions} lead to a linear system $KU=F$ where $K$ is the stiffness matrix, $U$ the displacements at the nodes and $F$ the forces applied on nodes. The general solution of this system can be decomposed as a linear combination of elemental solutions computed for each `pilot node' elemetal displacement. During the intervention, once the `pilot nodes' positions are retrieved, the global tissue deformation can be computed in real-time, as long as the initial stiffness matrix $K$ is not modified. 

Cyst drainage and tissue resection are repercuted on the stiffness matrix matrix by eliminating from $K$ the contributions of the elements modeling the cyst or the tumor. Once the cyst drainage is complete, all the cyst elements are removed from $K$ and every time a bit of tumoral tissue is resected, the element containing the surgical tool is identified and its contributions removed. Each such topological modification of the mesh requires a new general solution computation. The computational time of this operation is critical during the intervention.

\begin{figure}[htb]
\begin{minipage}[b]{1.0\linewidth}
  \centering \centerline{\epsfig{figure=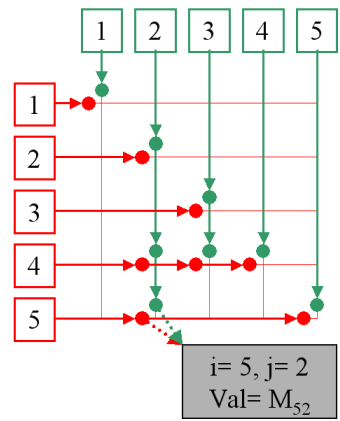,width=4.5cm}}
\end{minipage}
\caption{Data structure for sparse matrix. Each row and column is defined as a linked list of pointers at the matrix non-nul cells.}
\label{FigSparseMatrix}
\end{figure}
\begin{figure}[htb]
\begin{minipage}[b]{1.0\linewidth}
  \centering \centerline{\epsfig{figure=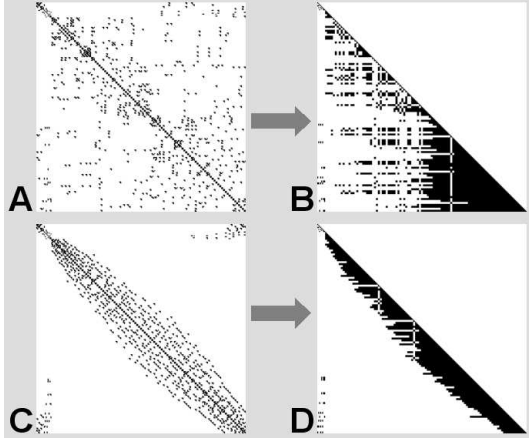,width=7cm}}
\end{minipage}
\caption{Sparse symmetric matrix (A) and its $LL^t$ triangular factor (B). The same initial sparse matrix after nodes reordering (C) and its triangular factor with reduced fill-in (D).}
\label{FigFillIn}
\end{figure}

The $KU=F$ system is solved using the $LL^t$ decomposition of the sparse symmetric positive-definite matrix $K$. This decomposition is optimised using the data structure dedicated to sparse matrices depicted in Figure \ref{FigSparseMatrix} as well as proper matrix conditioning. Figure \ref{FigFillIn} shows the impact of nodes ordering on the `fill-in' within the inferior triangular matrix $L$. A simple node ordering groups all non-null elements in $K$ near the diagonal. As a result, the $L$ matrix is much emptier and the system solution much faster. Finding the node ordering that minimises the `fill-in' of $L$ is a NP-complete problem. Algorithms such as Minimum Degree Ordering or its simpler version proposed by Amestoy et al. in \cite{Amestoy96} strongly reduce the number of non-null elements in $L$. Finally, the modification of the $J_0$ column in the $K$ matrix only affects the columns $J \geq J_0$ in the $L$ matrix thus it is a natural optimisation to put all the nodes forming part of elements likely to be resected at the `bottom-right' of the matrix by assigning them the highest indices.

\section{RESULTS}
\label{SecExampleOfSimulation}
\begin{figure}[htb]
\begin{minipage}[b]{1.0\linewidth}
  \centering \centerline{\epsfig{figure=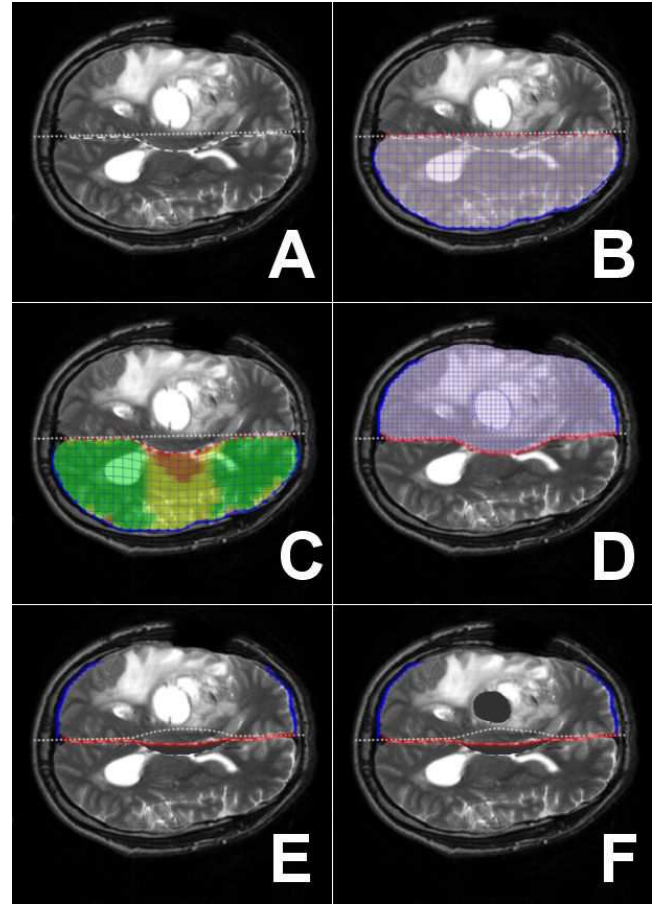,width=8.5cm}}
\end{minipage}
\caption{2D simulation of brain bulging and sagging.}
\label{FigSimulation}
\end{figure}

The upper hemisphere 2D mesh represented in Figure \ref{FigSimulation}-D has 1000 elements (mostly quadrilaterals). The deformation updates based on `pilot nodes' positions take about 3 miliseconds, while the resection updates require about 150 miliseconds. Implementation in 3D is still being carried out and in order to predict the performances we used a 2D mesh with a computational size similar to that of a 3D hexahedral elements with 5000 elements, among which 500 are dedicated to model the tumor. The computation delays were of about 400 miliseconds for deformation update and about 2300 miliseconds for resection update. 

Figure \ref{FigSimulation} depicts a 2D simulation of the initial brain-shift at dura opening and subsequent brain sagging after cyst and tissue resection. In this case, intracranial pressure excess creates an initial bulging, although this behaviour is not systematically observed.

Figure \ref{FigSimulation}-A: This preoperative T2 MRI scan shows the conflict between the two hemispheres. The dashed line shows the actual hemispheres interface while the dotted line indicates the initial interface position i.e. the middle of the skull. 

Figure \ref{FigSimulation}-B: Mesh of the contralateral hemisphere. The red nodes are `pilot' and the border blue nodes are `sliding'.

Figure \ref{FigSimulation}-C: Compression of the contralateral hemisphere. The color code displays the stress within the elements.

Figure \ref{FigSimulation}-D: Mesh of the hemisphere with tumor.

Figure \ref{FigSimulation}-E: Right after the dura opening, the stress computed in step C is applied to the upper mesh, which causes the brain to bulge.

Figure \ref{FigSimulation}-F: Gravity, cyst drainage and tissue resection cause the brain to sag.

\section{CONCLUSION} 
We proposed a general framework for a image-guided model-updated neuronavigator for tumor resection. This system is based upon ideas generally accepted by the community. Computation optimisations as well as use of appropriate data structures make it possible to model global parenchyma deformation as well as cyst drainage or tissue resection within time delays compatible with the intervention timing. 

From the biomechanical point of view, although some authors like Miller et al. in \cite{Miller00} propose complex hyperelastic models, we think that given the small deformation measured within the brain tissues the linear mechanical model may be acceptable. The small displacements hypothesis might be unadapted but a large displacement approximation based upon a series of elemental linear displacement can be the solution as suggested by Platenik et al. in \cite{Platenik02}.

Finally the 3D finite elements code implementing those features needs to be tested on phantom or patient data. Many clinical issues remain unresolved, such as the definition of the reference structures used to `pilot' the model as well as their intraoperative identification within the echographic planes.





\bibliographystyle{plain}
\bibliography{MyBib}

\begin{thebibliography}{10}

\bibitem{Audette03}
Audette~M A, Siddiqi K, Ferrie~F P, , and Peters~T M.
\newblock An integrated range-sensing, segmentation and registration framework
  for the characterization of intra-surgical brain deformations in image-guided
  surgery.
\newblock {\em Computer Vision and Image Understanding}, 89:226--251, 2003.

\bibitem{Hagemann99}
Hagemann A, Rohr K, Stiel~H S, Spetzger U, and Gilsbach~J M.
\newblock Biomechanical modeling of the human head for physically based
  non-rigid image registration.
\newblock {\em IEEE Transactions on Medical Imaging}, 18(10):875--884, 1999.

\bibitem{Nabavi01}
Nabavi A, Black P, Gering D, Westin C, Mehta V, Pergolizzi R, Ferrant M,
  Warfield S, Hata N, Schwartz R, Wells~III W, Kikinis R, and Jolesz F.
\newblock Serial intraoperative magnetic resonance imaging of brain shift.
\newblock {\em Neurosurgery}, 48(4):787--797, 2001.

\bibitem{Coenen05}
Volker~A Coenen, Timo Krings, Jurgen Weidemann, Franz-Joseph Hans, Peter
  Reinacher, Joachim~M Gilsbach, and Veit Rohde.
\newblock Sequential visualization of brain and fiber tract deformation during
  intracranial surgery with three-dimensional ultrasound: An approach to
  evaluate the effect of brain shift.
\newblock {\em Operative Neurosurgery}, 56(1):133--139, 2005.

\bibitem{Miga05}
Miga~M I, Sinha~T K, and Cash~D M.
\newblock Techniques to correct for soft tissue deformations during
  image-guided brain surgery.
\newblock {\em Payan Y., editor. Biomechanics Applied to Computer Assisted
  Surgery. Research Signpost Publisher, ISBN 81-308-0031-4}, pages 153--176,
  2005.

\bibitem{Miga03}
Miga~M I, Sinha~T K, Cash~D M, Galloway~R L, and Weil~R J.
\newblock Cortical surface registration for image-guided neurosurgery using
  laser-range scanning.
\newblock {\em IEEE Transactions on Medical Imaging}, 22(8):973--985, 2003.

\bibitem{Kyriacou99}
Kyriacou~S K, Davatzikos C, Zinreich~S J, and Bryan~R N.
\newblock Nonlinear elastic registration of brain images with tumor pathology
  using a biomechanical model.
\newblock {\em IEEE Transactions on Medical Imaging}, 18(7):580--592, 1999.

\bibitem{Bucki05}
Bucki M. and Payan Y.
\newblock Automatic finite elements mesh generation from planar contours of the
  brain: an image driven, 'blobby' approach.
\newblock {\em Payan Y., editor. Biomechanics Applied to Computer Assisted
  Surgery. Research Signpost Publisher, ISBN 81-308-0031-4}, pages 209--224,
  2005.

\bibitem{Comeau00}
Comeau~R M, Sadikot~A F, Fenster A, and Peters~T M.
\newblock Intraoperative ultrasound for guidance and tissue shift correction in
  image-guided neurosurgery.
\newblock {\em Medical Physics}, 27:787--800, 2000.

\bibitem{Ferrant02}
Ferrant M, Nabavi A, Macq B, Black~P M, Jolesz F, Kikinis R, and Warfield S.
\newblock Serial registration of intraoperative mr images of the brain.
\newblock {\em Medical Image Analysis}, 6:337--359, 2002.

\bibitem{Miller02}
K~Miller.
\newblock Biomechanics of brain for computer integrated surgery.
\newblock {\em Warsaw University of Technology Publishing House}, 2002.

\bibitem{Miller00}
K~Miller, K~Chinzei, G~Orssengo, and P~Bednarz.
\newblock Mechanical properties of brain tissue in-vivo: experiment and
  computer simulation.
\newblock {\em Journal of Biomechanics}, 33:1369--1376, 2000.

\bibitem{Hata00}
Hata N, Nabavi A, Wells~III W, Warfield S, Kikinis R, Black~P M, and Jolesz F.
\newblock Three-dimensional optical flow method for measurement of volumetric
  brain deformation from intraoperative mr images.
\newblock {\em Journal of Computer Assisted Tomography}, 24:531--538, 2000.

\bibitem{Clatz05}
Clatz O, Delingette H, Talos~I F, Golby~A J, Kikinis R, Jolesz~F A, Ayache N,
  and Warfield~S K.
\newblock Robust non-rigid registration to capture brain shift from
  intra-operative mri.
\newblock {\em Accepted for publication in IEEE TMI}, 2005.

\bibitem{Clatz05-RR}
Clatz O, Bondiau P, Delingette H, Sermesant M, Warfield S, Malandain G, and
  Ayache N.
\newblock Brain tumor growth simulation.
\newblock {\em INRIA, Research report 5187}, 2004.

\bibitem{Hastreiter00}
Hastreiter P, Reszk-Salama C, Nimsky C, Lurig C, Greiner G, and Ertl T.
\newblock Registration techniques for the analysis of the brain shift in
  neurosurgery.
\newblock {\em Computers and Graphics}, pages 385--389, 2000.

\bibitem{Hastreiter04}
Hastreiter P, Rezk-Salama C, Soza G, Bauer M, Greiner G, Fahlbusch R, Ganslandt
  O, and Nimsky C.
\newblock Strategies for brain shift evaluation.
\newblock {\em Medical Image Analysis}, 8:447--464, 2004.

\bibitem{Picinbono02}
G~Picinbono, H~Delingette, and N~Ayache.
\newblock Modèle déformable élastique non linéaire pour la simulation de
  chirurgie en temps réel.
\newblock {\em C. R. Biologies}, 325:335–344, 2002.

\bibitem{Platenik02}
L~A Platenik, M~I Miga, D~W Roberts, K~E Lunn, F~E Kennedy, A~Hartov, and K~D
  Paulsen.
\newblock In vivo quantification of retraction deformation modeling for updated
  image-guidance during neurosurgery.
\newblock {\em IEEE Transactions on Biomedical Eengineering}, 49(8):823--835,
  2002.

\bibitem{Amestoy96}
Amestoy~P R, Davis~T A, and Duff~I S.
\newblock An approximate minimum degree ordering algorithm.
\newblock {\em SIAM Journal on Matrix Analysis and Applications}, 17:886--905,
  1996.

\bibitem{Shattuck02}
Shattuck~D W and Leahy~R M.
\newblock Brainsuite: An automated cortical surface identification tool.
\newblock {\em Medical Image Analysis}, 6:129--142, 2002.

\end{thebibliography}

\end{document}